\documentclass[11pt]{article}
\usepackage[letterpaper,margin=1in,headheight=14pt]{geometry}
\usepackage{amsmath,amssymb}
\usepackage{iftex}
\ifPDFTeX\usepackage[T1]{fontenc}\fi
\usepackage{libertine}
\usepackage{booktabs}
\usepackage{array}
\usepackage{graphicx}
\usepackage[authoryear,round]{natbib}
\usepackage[hidelinks]{hyperref}
\usepackage{fancyhdr}
\usepackage{caption}
\usepackage{titlesec}
\usepackage{xurl}

\titleformat{\section}{\Large\bfseries}{\thesection}{1em}{}
\titleformat{\subsection}{\large\bfseries}{\thesubsection}{1em}{}
\graphicspath{{figures/}}

\fancypagestyle{plain}{\fancyhf{}\fancyfoot[C]{\thepage}}

\hypersetup{pdftitle={Evaluating Large Language Model Research Agents as Tests of Return Predictability}}

\title{Evaluating Large Language Model Research Agents as Tests of Return Predictability}
\author{Chang Liu\\[2pt]\small Department of Information Engineering and Computer Science, University of Trento, Italy}
\date{}

\begin{document}
\maketitle
\thispagestyle{plain}
\vspace{-1.5em}

\begin{abstract}
\noindent How often do research agents built on large language models report that they can forecast returns when the
data contain no signal? We score the conclusions of four such agents as statistical decisions on the same simulated
monthly return series. Without a prescribed significance level, the agents reported that their forecasts would beat
the expanding historical mean in 16--44\% of 110 null samples, whereas a Holm-adjusted regression test rejected the
absence of predictability in 8\% of the same samples. When a weak predictor was present, comparisons at matched
false-positive rates did not establish a detection advantage for the three commercial agents. Instructing those
agents to report outperformance only if they would reject the absence of predictability at the 5\% level, accounting
for every specification they evaluated, reduced their false-positive rates to 0--3\%. It also reduced detection at a
population $R^2$ of 1\%: from 74\% to 45\% for Claude Opus 5.5, from 59\% to 6\% for GPT-6.1 Sol and from 71\% to
15\% for GPT-6.1 Terra. Whether the instruction holds the false-positive rate at 5\% remains unresolved. For
researchers who delegate the search for return predictors to an agent, specifying the final comparison deserves as
much attention as choosing the forecasting model.

\end{abstract}


\section{Introduction}

An autonomous research agent built on a large language model can choose a forecasting model and then judge whether its forecast will beat a
benchmark. How much weight should a researcher give that judgment? In a simulation we can ask how often the agent
reports outperformance when returns contain no signal, and how often it does so when a predictor is present.

A procedure can detect more true signals by accepting weaker evidence, at the cost of more false positives.
With a conventional test, the researcher sets a significance level and can study the power of the resulting
procedure. An agent may choose both the specifications it examines and the evidence it requires before reporting that a model will outperform. Its verdict is then a test of unknown size and power.

In studies of return predictability, apparent gains over the historical mean
often disappear out of sample \citep{WelchGoyal2008, GoyalWelchZafirov2024}. Persistent regressors and specification
searches introduce further inferential difficulties \citep{Stambaugh1999, FersonSarkissianSimin2003,
HarveyLiuZhu2016}. Model-search corrections and out-of-sample tests are therefore central to evaluating the evidence
for a forecasting relation \citep{White2000, ClarkWest2007, CampbellThompson2008}. For agents that develop and
interpret their own analyses, the question is whether an equally explicit decision procedure remains in place.

Recent research shows variation in the findings of AI analysts given identical data \citep{GaoXiao2026,
BertranFogliatoWu2026} and examines autonomous pipelines for empirical finance \citep{QuChenWang2026, Gencay2026,
PanDingGiesecke2026}. Some of these workflows assign model discovery to an agent and evaluation to an independent procedure.
We study the case in which the agent also makes the final judgment, and we score that judgment against known
truth: how often it reports outperformance when no predictor exists, and how often it detects a predictor when its
false-positive rate is matched to that of a conventional test.

We use a Monte Carlo design adapted from studies of automatic model selection
\citep{CastleDoornikHendry2011,HendryDoornik2014}. Each agent sees 700 months of returns and eight candidate predictors. It can submit a model and
report whether it expects to beat the expanding historical mean. A separate evaluator scores the submitted forecast
over the next 360 months. The main experiment uses 350 samples with Gaussian innovations, each with at most one true linear predictor.
This design lets us compare the agent's conclusions with a conventional regression test under known signal strengths.

All four agents reported outperformance more often under the null than the conventional test rejected it. Once
false-positive rates were matched, none of the three commercial agents showed a clear detection advantage, and the
open-weight agent detected less often than the test. In a second experiment, a single sentence specifying a 5\% significance level reduced both false positives and
detection for the three commercial agents.

Section~2 describes the task, the agents and the simulated forecasting problems, and defines the regression test
against which their conclusions are compared. Section~3 reports how often the agents reach a positive conclusion with
and without a true predictor, before and after matching false-positive rates. Section~4 tests whether an explicit 5\%
decision rule changes these rates. Section~5 examines whether conclusions depend on a sentence about prior research,
how often repeated runs agree, and whether the agents' stated probabilities add information about their models'
out-of-sample performance beyond the conclusion itself. Section~6 discusses an alternative in which the comparison
rule is applied outside the agent, the limits of the design and the implications for forecasters.

\section{Experimental design}

\subsection{Task, agents and forecast outcomes}

Each run supplies monthly returns and eight anonymized predictor series. The instructions ask the agent to build a
forecasting model that it believes will outperform the expanding historical mean out of sample, while permitting a
conclusion that none will do so. Its final verdict is one of three: the model will outperform the benchmark, it will not, or no predictability was
found and the agent stops. We call the first a positive conclusion. The instructions specify mean
squared forecast error as the evaluation metric. Under the neutral instructions, agents choose how much evidence
to require before reporting outperformance.

The agents work in a network-isolated Python environment. They record a prior conclusion before inspecting data,
execute code, log the specifications they evaluate and submit a final verdict. They have a combined budget of 40
code and specification calls. Predictor values undergo random affine transformations, including possible sign
reversal; return units are randomly rescaled. These transformations reduce recognition of the underlying series. The three commercial agents, Claude Opus 5.5, GPT-6.1 Sol and GPT-6.1 Terra (hereafter Opus, Sol and Terra), use
their developers' high-reasoning settings. The open-weight Qwen3-32B model \citep{Yang2025Qwen3} runs locally through vLLM \citep{Kwon2023vLLM}.
These are different systems, and ``high'' reasoning effort is not the same setting across developers. In addition,
OpenAI supplied Sol and Terra only under undated aliases, so the underlying models may change over time.

The agents see 700 training observations. A separate evaluator assesses submitted models on 360 later observations
that the agents cannot access, refitting each model every 12 months. A submitted model beats the benchmark if
its out-of-sample $R^2$ is positive relative to the expanding mean. Even a genuine weak predictor can lose to the
benchmark over 360 months. The evaluator scores an available model
regardless of the agent's verdict. In the calibration analysis, runs without an evaluable forecast count as
not outperforming. Supplement S1 also reports results restricted to evaluable forecasts.

\subsection{Simulated forecasting problems}\label{sec:dgp}

The data-generating system is
\begin{equation}
r_{t+1} = \mu + \beta x_t + \varepsilon_{t+1}, \qquad x_{t+1} = \phi x_t + u_{t+1}, \qquad
\operatorname{corr}(\varepsilon_{t+1}, u_{t+1}) = \rho.
\end{equation}
The coefficient $\beta$ produces a population predictive-regression $R^2$ of 0, 0.25, 0.5 or 1\%. Predictor
persistence is $\phi \in \{0.5, 0.95\}$ and the innovation correlation is $\rho \in \{0, -0.8\}$. The latter permits
the finite-sample predictive-regression bias studied by \citet{Stambaugh1999}. Seven additional candidates are
autoregressive processes unrelated to returns, with persistence between 0.3 and 0.99. The true predictor, if present,
occupies a random position among the eight candidates.

The Gaussian experiment crosses the four $R^2$ values with both persistence and innovation-correlation settings,
giving 16 conditions and 20 independently generated samples per condition. Thirty more null samples contain a highly
persistent but nonpredictive candidate ($\phi = 0.98$, $\rho = -0.8$). There are consequently 350 distinct samples, including 110 null samples, of which 80 are from the linear design. All four agents receive the same samples. The 40
samples with $R^2 \in \{0, 1\%\}$, $\phi = 0.95$ and $\rho = -0.8$ are each run two more times, giving 430 runs per agent.
Analyses of conclusion rates use only the first run of a sample unless stated otherwise.

An earlier group of experiments drew innovations from a stationary bootstrap of residuals from the high-minus-low (HML) value factor
\citep{PolitisRomano1994, FamaFrench2015}. We use these samples for the paired experiment on prior-evidence wording.
Some submitted forecasting models matched sequences of repeated returns from the finite residual pool and used them to forecast the next
return. The primary error-rate estimates therefore come from the Gaussian samples. Supplement S2 describes the
bootstrap results. Table~\ref{tab:runs} lists the experiments.

\begin{table}[h]\centering
\caption{Experiments reported in this article}\label{tab:runs}
\small
\begin{tabular}{l l >{\raggedright\arraybackslash}p{6.4cm} r}
\toprule
Innovations & Models & Instructions and role & Runs \\
\midrule
Gaussian & Opus, Sol, Terra & Neutral; main conclusion-rate comparisons & 1,290 \\
Gaussian & Qwen3-32B & Neutral; open-weight comparison & 430 \\
Bootstrap & Opus & Neutral, optimistic, skeptical; instruction experiment and supporting evidence & 900 \\
Bootstrap & Sol, Terra & Neutral, optimistic, skeptical; instruction experiment and supporting evidence & 1,800 \\
Gaussian & Opus, Sol, Terra & Explicit decision rule, with renewed neutral controls for Sol and Terra & 950 \\
\bottomrule
\end{tabular}

\smallskip
\parbox{0.85\linewidth}{\footnotesize Innovations: distribution of the shocks in the simulated samples (Section~\ref{sec:dgp}).
Neutral: the task instructions alone. Optimistic and skeptical: the same instructions with one appended sentence on
prior evidence (Section~\ref{sec:framing}). Explicit decision rule: the instructions with the appended 5\% sentence
(Section~\ref{sec:rule}). Runs: all agent runs, including repeated runs on the same sample. In the Gaussian
experiments each agent has 430 runs (350 samples, 40 of them run three times); in the decision-rule experiment each
commercial agent has 190 runs under the rule (110 null and 80 signal samples), and Sol and Terra have 190 new neutral
runs each.}

\end{table}

\subsection{Comparison procedure and inference}\label{sec:inference}

The comparison procedure tests each of the eight supplied predictors and the mean of the asset's monthly returns
over the preceding 12 months. Each training-sample regression uses a two-sided $t$ test with Newey--West heteroskedasticity- and autocorrelation-consistent (HAC) standard errors at 12 lags \citep{NeweyWest1987}. Holm's step-down correction adjusts for the nine tests \citep{Holm1979}.
We call this procedure the HAC--Holm test. It tests for a predictive relationship, while the agent judges whether
its own model will beat the benchmark out of sample.

We score both decisions against the known presence or absence of a predictor. A positive decision under the null
counts as a false positive. A positive decision when a predictor exists counts as a detection, regardless of which
variable the agent selected or how its forecast performed in the evaluation period. For the regression test these
rates estimate size and power.

In an exploratory comparison, we adjust the test's nominal level to match each agent's observed false-positive
rate in the 80 linear null samples. The simulated calibration sets the level using 8,000 independent simulated null samples
and estimates the test's detection rate on independently simulated signal samples. The same-sample calibration uses
the null and signal samples the agents analyzed for these two steps.

We report Wilson 95\% intervals for rates. Matched comparisons use exact McNemar tests, and experiments with three
paired instructions use Cochran's $Q$. Size-adjusted intervals use 2,000 resamples that recalibrate the test in each
draw. The simulated sample, not an individual repeated run, is the resampling unit. The Gaussian false-positive tests and paired null comparisons were specified before the runs; size-adjusted power,
model-outcome calibration and the variance decomposition of repeated runs are exploratory. Appendix G of the
Supplement lists the prespecified hypotheses, and Appendix H the analyses designed after the results were known.

\section{How reliably do agents identify predictability?}

\subsection{Positive conclusions when no predictor exists}\label{sec:nullrates}

In the 110 Gaussian null samples, Opus reports outperformance in 37 cases (34\%, Wilson 95\% interval 25--43\%), Sol
in 18 (16\%), Terra in 48 (44\%) and Qwen3-32B in 39 (35\%). The HAC--Holm test rejects in 8\% of these same samples at
its nominal 5\% level (Table~\ref{tab:main}). Every agent reports more positive conclusions than the test in the
prespecified paired comparisons (Holm-adjusted $p \le 0.006$). Each agent's own rate also exceeds 5\% in the
prespecified one-sided binomial tests (adjusted $p < 0.001$). The three commercial agents differ in their null-sample
conclusion rates (Cochran's $Q = 28.8$, adjusted $p < 0.001$).

The rates are similar in the 80 linear null samples used for the size adjustment below: 36\% for Opus, 16\% for
Sol, 42\% for Terra and 40\% for Qwen3-32B. Because the neutral instructions set no significance level, these rates
describe the agents' implicit evidence thresholds, not departures from a requested one.

\begin{table}[h]\centering
\caption{Positive-conclusion rates and comparisons with the HAC--Holm test}\label{tab:main}
\small
\begin{tabular}{l rr rr rr}
\toprule
& \multicolumn{2}{c}{Null (110)} & \multicolumn{2}{c}{$R^2 = 1\%$ (80)} & \multicolumn{2}{c}{Size-adjusted difference} \\
\cmidrule(lr){2-3}\cmidrule(lr){4-5}\cmidrule(lr){6-7}
Model & Agent & HAC--Holm & Agent & HAC--Holm & Simulated & Same-sample \\
\midrule
Opus & 34 [25,43] & 8 & 74 & 56 & $-10$ [$-21$,2] & 6 [$-6$,14] \\
Sol & 16 [11,24] & 8 & 57 & 56 & $-10$ [$-23$,8] & $-1$ [$-15$,12] \\
Terra & 44 [35,53] & 8 & 84 & 56 & $-3$ [$-13$,7] & 11 [$-2$,24] \\
Qwen3-32B & 35 [27,45] & 8 & 44 & 56 & $-42$ [$-53$,$-29$] & $-26$ [$-45$,$-10$] \\
\bottomrule
\end{tabular}

\smallskip
\parbox{0.9\linewidth}{\footnotesize Entries are percentages, except the size-adjusted differences, which are
percentage points. Null (110): share of the 110 null samples in which the agent reports outperformance (its
false-positive rate) and in which the HAC--Holm test rejects at nominal 5\%. $R^2 = 1\%$ (80): the same shares in the
80 samples whose true predictor has population $R^2 = 1\%$ (detection rates). Size-adjusted difference: the agent's
detection rate at $R^2 = 1\%$ minus that of a HAC--Holm test whose level is set to reproduce the agent's
false-positive rate in the 80 linear null samples (36, 16, 42 and 40\%; Section~\ref{sec:nullrates}). A positive
value means that the agent detects a predictor more often than an equally permissive test. Simulated: the test's
level and detection rate are estimated on independently simulated samples. Same-sample: both are estimated on the
samples the agents analyzed. Brackets: Wilson 95\% intervals for the agents' null rates and bootstrap 95\% intervals
for the size-adjusted differences, which are exploratory.}

\end{table}

\subsection{Positive conclusions when a predictor is present}

\begin{figure}[t]\centering
\includegraphics[width=0.72\linewidth]{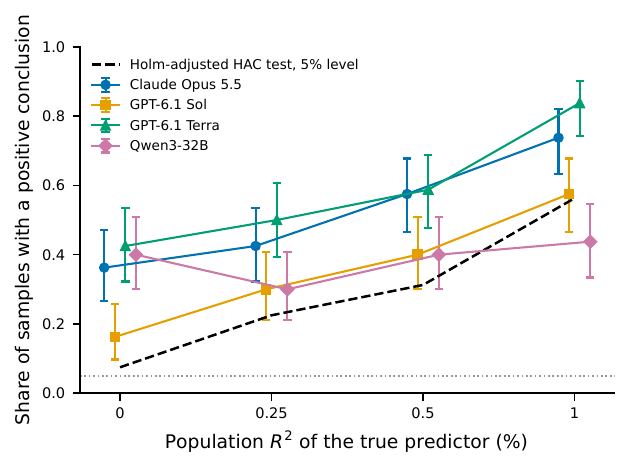}
\caption{Share of Gaussian linear samples receiving a positive conclusion at each population $R^2$ (80 samples per
level, first runs). Error bars indicate Wilson 95\% intervals. The dashed line is the HAC--Holm test at nominal 5\%
and the dotted line marks 5\%.}\label{fig:rates}
\end{figure}

For the commercial agents, positive conclusions become more common as the signal grows (Figure~\ref{fig:rates}). At
a population $R^2$ of 1\%, Opus concludes that its model beats the benchmark in 74\% of samples, Sol in 57\% and Terra
in 84\%. The training-sample HAC--Holm test detects a predictor in 56\% of these samples; it does not directly test
whether a particular submitted forecast wins in the holdout. Qwen3-32B changes little between the null and the
strongest signal: its rates are 40\% and 44\% in the 80-sample linear conditions. Higher unadjusted detection by Opus
and Terra is not evidence of a better search: those agents also report outperformance much more often under the null.

\begin{figure}[t]\centering
\includegraphics[width=\linewidth]{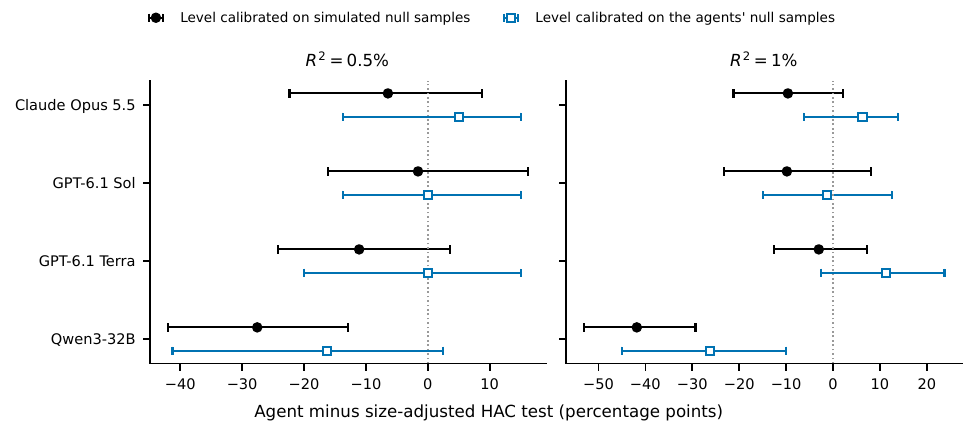}
\caption{Differences in detection rates after matching false-positive rates in the Gaussian design, at population $R^2$ of 0.5\% (left) and 1\% (right).
Values are percentage points (agent minus HAC--Holm test). Dots: simulated calibration on 8,000 independent null samples.
Open squares: same-sample calibration on the 80 linear null samples analyzed by the agents. Bars show bootstrap 95\% intervals for these
exploratory comparisons.}\label{fig:sizeadj}
\end{figure}

The size-adjusted comparison measures discrimination directly (Figure~\ref{fig:sizeadj}). Under the simulated
calibration, agent-minus-test differences at $R^2 = 1\%$ are $-10$
percentage points for Opus (95\% interval $-21$ to 2), $-10$ for Sol ($-23$ to 8) and $-3$ for Terra ($-13$ to 7).
The same-sample calibration gives 6 ($-6$ to 14), $-1$ ($-15$ to 12) and 11
($-2$ to 24). Each interval includes zero. For Terra, for example, the same-sample interval extends from a 2-point disadvantage
to a 24-point advantage over the test. Qwen3-32B is below the test under both calibrations, with differences of $-42$ and $-26$ points and intervals excluding zero. Because the design contains at most one
true linear predictor, which the test examines directly, it leaves agents little room to gain by constructing a
nonlinear forecast.

The two calibrations differ because the agents' null samples are unusual. In 41 of the 80 linear null samples
(51\%), the best supplied predictor has a Newey--West $t$ statistic significant at 5\% before the Holm adjustment, compared with 39\% in the larger simulated
null population. This concentration of misleading candidates changes the test's behavior at the more
permissive thresholds needed for size matching. The simulated calibration represents a typical null environment;
the same-sample calibration matches the environments the agents actually encountered but has greater sampling
uncertainty. We therefore report both.

\section{Does stating a decision rule change the conclusion?}\label{sec:rule}

The initial instructions allow the agent to decide what constitutes sufficient evidence. The decision-rule experiment
appends the following sentence to the otherwise unchanged task instruction:
\begin{quote}\small
``Answer beats only if you would reject the hypothesis of no out-of-sample predictability at the 5\% significance
level, taking into account every specification you evaluated.''
\end{quote}
The sentence names a threshold and acknowledges selection but prescribes no inferential test, so agents still
choose how to account for the specifications they examined.

The three commercial agents analyzed the 110 null samples and the 80 samples with $R^2 = 1\%$ under the rule. Opus's
original neutral runs serve as its control. Sol and Terra were rerun under neutral instructions interleaved with the
rule runs, because their model aliases might have changed. Those new neutral null rates, 20\% and 41\%, are close to
the original 16\% and 44\%. Table~\ref{tab:rule} compares each model with its own control.

\begin{table}[h]\centering
\caption{Effect of the explicit 5\% decision-rule instruction}\label{tab:rule}
\small
\begin{tabular}{l rr rr c}
\toprule
& \multicolumn{2}{c}{Null (110)} & \multicolumn{2}{c}{$R^2 = 1\%$ (80)} & Null stops \\
\cmidrule(lr){2-3}\cmidrule(lr){4-5}
Model & No rule & Rule & No rule & Rule & No rule / rule \\
\midrule
Opus & 34 [25,43] & 3 [1,8] & 74 [63,82] & 45 [35,56] & 32 / 3 \\
Sol & 20 [14,28] & 0 [0,3] & 59 [48,69] & 6 [3,14] & 25 / 0 \\
Terra & 41 [32,50] & 1 [0,5] & 71 [61,80] & 15 [9,24] & 54 / 43 \\
HAC--Holm test & \multicolumn{2}{c}{8} & \multicolumn{2}{c}{56} & --- \\
\bottomrule
\end{tabular}

\smallskip
\parbox{0.8\linewidth}{\footnotesize Null (110): share of the 110 null samples with a positive conclusion (the
false-positive rate), in percent. $R^2 = 1\%$ (80): share of the 80 samples whose true predictor has population
$R^2 = 1\%$ with a positive conclusion (the detection rate), in percent. No rule: neutral instructions. Rule: the same
instructions with the appended 5\% sentence. Opus's no-rule runs are those of the main experiment; Sol and Terra were
rerun under neutral instructions alongside their rule runs. Null stops: number of the 110 null samples in which the
agent stopped with an explicit no-predictability verdict, whether or not a model was available to score. Brackets:
Wilson 95\% intervals. HAC--Holm test: the unchanged comparison test at nominal 5\% on the same samples.}
\end{table}

The rule reduces null-sample positive conclusions from 34\% to 3\% for Opus, from 20\% to zero for Sol and from 41\%
to 1\% for Terra. Across the three sets of paired model--sample decisions, 100 switch from a positive conclusion to a
nonpositive one, with none switching the other way. The average decrease is 30 percentage points (cluster-resampled
interval 24--37). The decrease does not reflect runs ending without a verdict: every null-sample run returned one, and explicit
stop verdicts became less frequent (Table~\ref{tab:rule}). None of the three one-sided tests rejects a false-positive
rate of 5\% under the instruction in favor of a higher one. For Opus, however, the Wilson interval extends to 8\%, so
a population rate above the requested level remains possible.

Detection also declines. At $R^2 = 1\%$, Opus falls from 74\% to 45\%, Sol from 59\% to 6\% and Terra from 71\% to
15\%. Two exploratory benchmarks put these rates in context; both are descriptive, not size-matched tests. A HAC--Holm
test calibrated to Opus's observed 3.8\% false-positive rate in the linear null samples detects about 39\% of
simulated signal samples, against 45\% for Opus. Sol and Terra report no positive conclusion in those 80 samples, so
their rates cannot be matched; a HAC--Holm test calibrated to a 1\% null rate detects about 25\%, against 6\% for Sol
and 15\% for Terra. The detection cost of the instruction therefore depends on the model and cannot be inferred from
its false-positive rate under neutral instructions.

\section{Further evidence on the decisions}

\subsection{Dependence on stated prior evidence}\label{sec:framing}

\begin{figure}[t]\centering
\includegraphics[width=\linewidth]{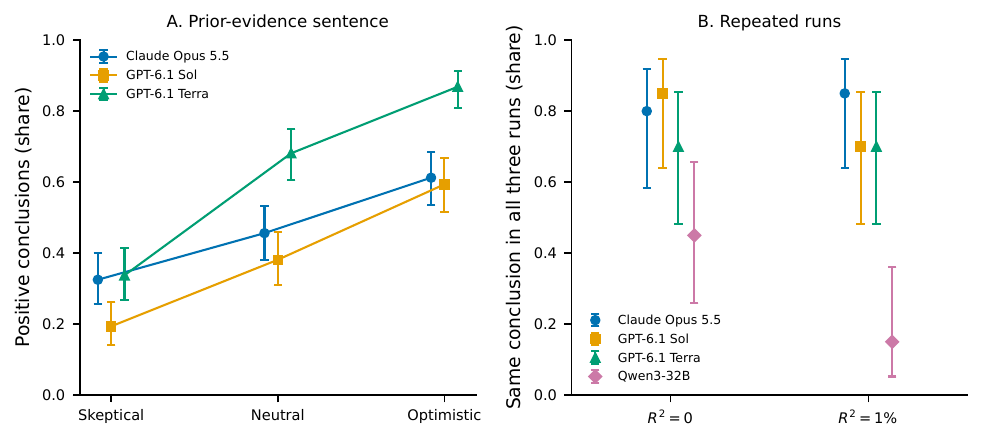}
\caption{Positive conclusions under three descriptions of prior evidence (left, 160 paired bootstrap samples)
and agreement across three runs on each Gaussian sample (right, 20 samples per $R^2$ level). The right panel
counts a sample as an agreement when all three binary decisions match. Bars show Wilson 95\% intervals.}\label{fig:framing}
\end{figure}

A separate paired experiment gives each of the three commercial agents identical bootstrap samples under three
instructions: neutral, an appended sentence saying prior research indicates a genuine timing signal, or an appended
sentence saying timing is very unlikely to work. For 160 linear samples with population $R^2$ of 0, 0.5, 1 or 2\%,
the conclusion changes substantially with that single sentence (Figure~\ref{fig:framing}). Opus's positive-conclusion
rate is 32\% under skeptical instructions, 46\% under neutral and 61\% under optimistic instructions. The
corresponding rates are 19, 38 and 59\% for Sol, and 34, 68 and 87\% for Terra. Cochran's $Q$ rejects invariance to
the instruction for each model (Holm-adjusted $p < 0.001$), as prespecified.

The direction is not confined to genuinely predictive samples: the optimistic sentence also raises null-sample
positive conclusions. Some forecasting models submitted by Terra exploit duplicated innovations in the bootstrap design, and the number
of exploiting runs varies by instruction: 17 under optimistic, 19 under neutral and five under skeptical
instructions. Excluding them would condition on an outcome affected by the instruction. We instead use a
least-favorable recoding: every identified exploiting run is treated as a nonpositive conclusion under optimistic
instructions and a positive conclusion under skeptical instructions. Even then, Terra's optimistic-minus-skeptical
difference is 42 percentage points (95\% interval 34--51), compared with 53 points without recoding. This check
supports the direction of the instruction effect but cannot repair the bootstrap design.

\subsection{Variation between repeated runs}

Repeated runs on exactly the same Gaussian samples provide a separate check on reproducibility. For 20 null samples
at the high-persistence design point, all three conclusions match in 80\% of cases for Opus, 85\% for Sol and 70\%
for Terra. At $R^2 = 1\%$ the proportions are 85, 70 and 70\%. Pooling the two conditions, the probability that two
runs disagree is 12, 15 and 20\%, respectively. In an exploratory decomposition, run-to-run variation on the
same sample accounts for 26\% of the within-condition variance of a single decision for Opus, 41\% for Sol and 46\%
for Terra.
For Qwen3-32B, all three runs agreed in 45\% of the null samples and 15\% of the samples with $R^2=1\%$.

\subsection{Stated probabilities and realized forecasting outcomes}

Agents also state a probability that their submitted model will beat the benchmark in the held-out period. This is
an assessment of realized model performance, not a probability that the data-generating process has a nonzero
coefficient. In all 3,990 commercial-agent runs of the main Gaussian and the bootstrap experiments (Table~\ref{tab:runs}), a positive conclusion is reported exactly when the stated probability is at least 0.5. Under the decision rule the same relationship holds in 189, 186
and 182 of 190 runs for Opus, Sol and Terra, respectively. The two outputs are therefore closely linked (Supplement S4).

In the Gaussian samples, the submitted model beats the benchmark after 41\% of Opus's positive conclusions, 48\% of
Sol's, 35\% of Terra's and 8\% of Qwen3-32B's, whereas the mean stated probabilities attached to those conclusions are
59, 69, 66 and 54\%. Adding the stated probability to a prediction based on the binary conclusion alone reduces the
five-fold cross-validated Brier score by at most 0.005, and all bootstrap intervals for the change include zero. In
this exploratory analysis, the probability adds little to predicting which models will beat the historical mean over
360 months. Supplement Figure S1 and Appendix D show how these results depend on the treatment of runs without an
evaluable model.

\section{Discussion and conclusion}

Scored as statistical decisions, the agents' verdicts had high and model-specific false-positive rates. Under
neutral instructions they reported outperformance in 16--44\% of null samples, against 8\% for a Holm-adjusted
regression test, and once false-positive rates were matched none showed a clear advantage in detecting a weak
predictor.

In these experiments, the same agent chose which forecasting models to try and how much evidence justified a
positive conclusion. A researcher could instead ask the agent to document its search and submit a forecast, then
apply a prespecified comparison rule to observations withheld from model development. This makes the criterion
explicit and allows its error rate to be studied. In an exploratory application to the Gaussian samples, a
one-sided Clark--West test of the submitted forecasts in the held-out period gave no more than 5\% positive
conclusions under the null but detected only 25--28\% of commercial-agent samples at $R^2 = 1\%$. Such a rule is
valid only if the withheld observations are independent of those used for development. In the bootstrap samples,
return sequences recurred in both periods, allowing some submitted models to exploit the same pattern in the
held-out observations.

The principal data-generating process contains no more than one supplied linear predictor. That restriction allows
comparisons with conventional regression tests but may understate the value of constructing more complex forecasts.
The 80 linear null samples also contain unusually many spuriously significant candidates, which helps explain
why the two size calibrations disagree. The sample includes four agents under one tool
configuration and budget, with two commercial models accessed through undated aliases. Withholding the evaluation
period was a fixed feature of the study, not an experimentally varied intervention. The decision-rule experiment
concerns only three commercial agents and two signal strengths, and its instruction specifies no test procedure. These results do not establish general error rates for
autonomous agents working on other research problems.

Consider a forecaster who uses the agent's model after a positive conclusion and the historical mean otherwise.
At a population $R^2$ of 1\% in the Gaussian design, this strategy's average out-of-sample $R^2$ ranged from
0.16\% to 0.17\% for the commercial agents and was $-5.74\%$ for Qwen3-32B; the corresponding strategy based on
the HAC--Holm test achieved 0.23\%. All four agents' averages were negative in the linear null samples (Supplement,
Appendix D). Forecast claims on published series face a further difficulty: without tools, Opus could recall some
historical factor returns (Supplement S3).

For researchers using agents to search for return predictors, specifying the final comparison deserves as much
attention as choosing the forecasting model. Here, naming a significance level reduced false positives but also
reduced detection, and the agents retained discretion over how to conduct the test.

\section*{Data and code availability}
The replication package contains code and configuration, data hashes, complete agent transcripts and the analysis plans. Its scripts regenerate the reported tables and figures without new paid API calls.
Re-running the agents requires access to the commercial APIs or local GPU hardware, as appropriate. The package has
not yet been deposited in a public repository; deposition with a persistent identifier is planned upon acceptance.

\section*{Acknowledgments}
The author used Claude Code (Anthropic) and Codex (OpenAI) for language editing of the manuscript and supporting
information. The author takes responsibility for the content.

\clearpage
\renewcommand{\listfigurename}{Figure legends}
\listoffigures

\end{document}